\documentclass[pdftex,aps,tightenlines,showpacs]{revtex4}

\usepackage{float}		      
\usepackage{dsfont}
\usepackage{graphicx}            

	\newcommand{\ket}[1]{\ensuremath{|#1\rangle}}			
	\newcommand{\bra}[1]{\ensuremath{\langle#1|}}			
  \newcommand{\iprod}[2]{\ensuremath{\langle#1|#2\rangle}}	
  \newcommand{\oprod}[2]{\ensuremath{|#1\rangle\langle#2|}}	
\newcommand{\Tr}{\ensuremath{{\rm Tr}}}				
\begin{document}
\title{Geometric measures of entanglement and the Schmidt decomposition}
\author{D.~Ostapchuk}
\email{dcmostap@uwaterloo.ca}
\author{G.~Passante}
\email{glpassan@mail.iqc.uwaterloo.ca}
\author{R.~Kobes}
\email{r.kobes@uwinnipeg.ca}
\author{G.~Kunstatter}
\email{g.kunstatter@uwinnipeg.ca}
\affiliation{Physics Department \\and\\ Winnipeg Institute for Theoretical Physics\\
University of Winnipeg\\ Winnipeg, MB\ R3B 2E9\\Canada}
\begin{abstract}
In the standard geometric approach to a measure of entanglement of a pure state,
$\sin^2\theta$ is used, where $\theta$ is the angle between the state to the 
closest separable 
state of products of normalized qubit states. We consider here a generalization of
this notion to separable states consisting of 
products of unnormalized states of different dimension. In so doing, 
the entanglement measure $\sin^2\theta$ is found to have an interpretation as the
distance between the state to the closest separable state. We also
find the components of the closest separable state and its norm have
an interpretation in terms of, respectively, the eigenvectors and eigenvalues 
of the reduced density matrices arising in the Schmidt decomposition of 
the state vector.
\end{abstract}
\maketitle
\section{Introduction}
With recognition of its role as a resource in quantum computing \cite{nielsen},
the nature of entanglement in quantum systems
is a problem of much current interest \cite{review1,review2,review3}. 
Of particular importance is
a quantitative measure of entanglement \cite{quantify}; two of the more commonly 
used measures, depending on the context,
are the von Neumann entropy, based on reduced density matrices \cite{review1},
and a geometric measure, based on the distance to the nearest product
state \cite{g1,g2,g3,g4,witness}. 
\par
The use of the von Neumann entropy as a measure of entanglement can be
understood in the framework of the Schmidt decomposition 
\cite{nielsen,s1,s2,s3,s4,s5}.
Consider a pure state $\ket{\psi}$ in an $n$-dimensional Hilbert space,
assumed to be normalized: $\iprod{\psi}{\psi}=1$. Let us then decompose 
the system into an $u$-dimensional subsystem, $A$, and a $v$-dimensional
subsystem, $B$, such that $n = uv$. For a basis $\ket{i}$ of $A$ and
$\ket{j}$ of $B$ we can write $\ket{\psi}$ as
\begin{equation}
\ket{\psi} = \sum_{i=0}^{u-1}\sum_{j=0}^{v-1} \gamma_{ij} \ket{i}\otimes \ket{j}
\end{equation}
for some complex coefficients $\gamma_{ij}$. However, the Schmidt decomposition
states that there exists
a basis $\ket{\alpha_i}$ for $A$ and 
$\ket{\beta_j}$ for $B$ such that $\ket{\psi}$ can be expressed as
a single summation:
\begin{equation}
\ket{\psi} = \sum_{k=0}^{ {\rm min}(u-1, v-1) } \sqrt{p_k} \ 
\ket{\alpha_k} \otimes \ket{\beta_k} 
\label{sd}\end{equation}
where the Schmidt coefficients satisfy $\sum_k p_k = 1$. The existence of
this decomposition follows from the singular value decomposition of the
matrix of coefficients $\gamma_{ij}\ket{i}\otimes\bra{j}$, and can be
related to the 
reduced density matrices formed by tracing out one of the subsystems 
of the density matrix $\rho = \oprod{\psi}{\psi}$:
\begin{eqnarray}
\rho_A &=& \Tr_B(\rho) = \sum_k p_k \ \oprod{\alpha_k}{\alpha_k} \nonumber\\
\rho_B &=& \Tr_A(\rho) = \sum_k p_k \ \oprod{\beta_k}{\beta_k}
\end{eqnarray}
by which one can see that $\alpha_k$ are the eigenvectors of $\rho_A =\Tr_B(\rho)$
and $\beta_k$ are the eigenvectors of $\rho_B = \Tr_A(\rho)$. The spectrum of
eigenvalues of $\rho_A$ and $\rho_B$, which are the same, can be used to
quantify the degree of entanglement of $\ket{\psi}$ due to the fact that, for a separable
state, only one non--zero eigenvalue is present. In this context 
$K = 1/\sum_k p_k^2$,
which satisfies $K \ge 1$, is often considered; 
another commonly used measure is the von Neumann entropy:
\begin{equation}
S = -\Tr\left(\rho_A \ln \rho_A\right) = -\Tr\left(\rho_B \ln \rho_B\right)
= -\sum_k p_k \log_2 p_k
\end{equation}
\par
On the other hand, in the geometric approach to measuring entanglement, 
one considers the distance from a $n$--dimensional
pure state
$\ket{\psi} = \sum_{i=0}^{n-1}\chi_{p_i}\ket{e_{p_i}^{(i)}}$ 
to a separable state
\begin{equation}
\ket{\phi} = \otimes_{i=0}^{n-1}\ket{\phi^{(i)}}
 = \sum_{p_i} c_{p_i}^{(i)} \ket{e_{p_i}^{(i)}}
\end{equation}
where $i=0, 1, \ldots, n-1$. The states $\ket{\phi^{(i)}}$ here are
assumed to be normalized: $\iprod{\phi^{(i)}}{\phi^{(i)}} = 1$. 
Minimizing $|\ket{\phi} - \ket{\psi}|^2$ results in the non--linear
eigenvalue equations:
\begin{equation}
\sum_{p_0 \cdots {\widehat p_i}\cdots p_{n-1} } \chi^*_{p_0 \cdots p_{n-1}}
c_{p_0}^{(0)} \cdots \widehat {c_{p_i}^{(i)}} \cdots c_{p_{n-1}}^{(n-1)}  
= \Lambda c_{p_i}^{(i)*}
\label{normalized}\end{equation}
where the eigenvalue $\Lambda$ is the Lagrange multiplier enforcing 
$\iprod{\phi^{(i)}}{\phi^{(i)}} = 1$ and $\ {\widehat {}}\ $ means omission.
The eigenvalues $\Lambda$ can be shown to lie in the range $-1\le \Lambda \le 1$,
and so are interpreted as the cosine of the angle between $\ket{\phi}$ and
$\ket{\psi}$; this can be seen by multiplying Eq.~(\ref{normalized}) by
$c_{p_i}^{(i)}$ and summing over $i$. A measure of entanglement is then taken to be
$1 - \Lambda_{\rm max}^2$, where $\Lambda_{\rm max}$ 
corresponds to the eigenvalue determined from Eq.~(\ref{normalized})
of the closest separable state.
\par
In this paper we show that if one uses {\it unnormalized} separable states in
a geometric measure of entanglement, the norm of the closest separable state
can be related to both the distance between and to the angle
between the separable and target states; this will provide a natural interpretation
of the entanglement measure $1 - \Lambda_{\rm max}^2$ as
the distance to the closest separable state. As well, by considering
the geometric measure in arbitrary dimensional spaces, a connection can be established
between the basis states and eigenvalues of
the Schmidt decomposition of the state and the components and norm 
of the closest separable product state used in this geometric approach.
\section{Optimum Euclidean Distance}
In this section we describe the geometric measure of entanglement we shall use
based on finding the extrema of the distance to a separable state.
For this, we consider an $n$--dimensional
pure state
$\ket{\psi} = \sum_{i=0}^{n-1}\chi_{p_i}\ket{e_{p_i}^{(i)}}$.
 We split the $n$--dimensional
space into subspaces $A, B, C, \ldots$ of dimension $u, v, \ldots $, 
and consider the separable state
\begin{equation}
 \ket{\phi} = \ket{A}\otimes\ket{B} \otimes \ket{C}\otimes\ldots
   = \sum_{i=0}^{u-1} a_i\ket{e_A^{(i)}} \otimes 
   \sum_{j=0}^{v-1} b_j\ket{e_B^{(j)}} \otimes
   \sum_{k=0}^{w-1} c_j\ket{e_C^{(k)}} \otimes\ldots
\end{equation}
The state $\ket{\phi}$ is not assumed to
be normalized. 
We now form the distance from the state $\ket{\psi}$ to such a separable state:
\begin{equation}
D^2 = \left| \ket{\phi} - \ket{\psi}\right|^2
= \left( \sum_{i = 0}^{u-1}\sum_{j = 0}^{v-1}\sum_{k = 0}^{w-1}\cdots \right)
\left(a_i^*b_j^*c_k^*\ldots - \chi_{ijk\cdots}^*\right)
\left(a_ib_jc_k\ldots - \chi_{ijk\cdots}\right)
\end{equation}
where $a_i, b_j, \ldots$ are the 
coordinates in the appropriate spaces,
and optimize this distance with respect to the coordinates of 
$\ket{A}, \ket{B}, \ldots$:
\begin{eqnarray}
\frac{\partial D^2}{\partial a_i} = 0 
&\Rightarrow& a_i^* N_BN_C\ldots =
\left(\sum_{j=0}^{v-1} \sum_{k=0}^{w-1}\cdots \right)
 b_jc_k\ldots\chi_{ijk\ldots}^*\nonumber\\
\frac{\partial D^2}{\partial b_j} = 0 
&\Rightarrow& b_j^* N_AN_C\ldots =
\left(\sum_{i=0}^{u-1} \sum_{k=0}^{w-1}\cdots \right)
a_ic_k\ldots \chi_{ijk\ldots}^*\nonumber\\
\frac{\partial D^2}{\partial c_k} = 0 
&\Rightarrow& c_k^* N_AN_B\ldots =
\left(\sum_{i=0}^{u-1} \sum_{j=0}^{v-1}\cdots \right)
a_ib_j\ldots \chi_{ijk\ldots}^*\nonumber\\
&\vdots&\label{opt}\end{eqnarray}
where 
$N_A =\displaystyle \sum_{i = 0}^{u-1}a_i^*a_i$,
$N_B =\displaystyle \sum_{j = 0}^{v-1}b_j^*b_j$, and so on. Except in special
cases these non--linear equations must be solved numerically. However,
from Eq.(\ref{opt}), one can show
\begin{equation}
N_A N_BN_C\ldots =\left(\sum_{i=0}^{u-1}\sum_{j=0}^{v-1} \sum_{k=0}^{w-1}\cdots\right)
a_i b_jc_k\ldots\chi_{ijk\ldots}^*
\end{equation}
and hence, at the critical points,
\begin{equation}
\iprod{\phi}{\psi} = N_AN_BN_C\ldots = \iprod{\psi}{\phi}
\end{equation}
Using $\iprod{\psi}{\psi} = 1$ and $\iprod{\phi}{\phi} = N_AN_BN_C\ldots$, 
we find that, at the critical point, the angle between $\ket{\phi}$ and
$\ket{\psi}$ is
\begin{equation}
\cos\theta_C = \left.
\frac{ \iprod{\psi}{\phi}}{\sqrt{\iprod{\phi}{\phi}}\sqrt{\iprod{\psi}{\psi}}}
\right|_{\rm critical} =
\sqrt{N_AN_BN_C\ldots}
\label{cangle}\end{equation}
and the distance $D^2$ is
\begin{equation}
D_C^2 = 1-N_AN_BN_C\ldots = \sin^2\theta_C
\label{cdistance}\end{equation}
\par
Consistency of the above requires $N_AN_BN_C\ldots \le 1$.
This can be done using the Cauchy--Schwartz inequality:
\begin{equation}
\left| \iprod{\phi - \psi}{\phi} \right|^2 
\le \iprod{\phi - \psi}{\phi - \psi} \iprod{\phi}{\phi}
\label{cs}\end{equation}
Using $\iprod{\phi}{\phi} = N_AN_BN_C\cdots$ and 
the definition $D^2 = \iprod{\phi - \psi}{\phi - \psi}$, this then implies
\begin{equation}
N_AN_BN_C\cdots D^2 \ge \left| \iprod{\phi}{\phi} - \iprod{\psi}{\phi} \right|^2
\end{equation}
At the extremal points we have $\iprod{\psi}{\phi} = N_AN_BN_C\cdots$ and
$D^2 = 1-N_AN_BN_C\cdots$, and so we can conclude
\begin{equation}
1-N_AN_BN_C\cdots \ge 0 \Rightarrow N_AN_BN_C\cdots \le 1
\end{equation}
\par
\begin{center}
\begin{figure}
\includegraphics[width=220pt]{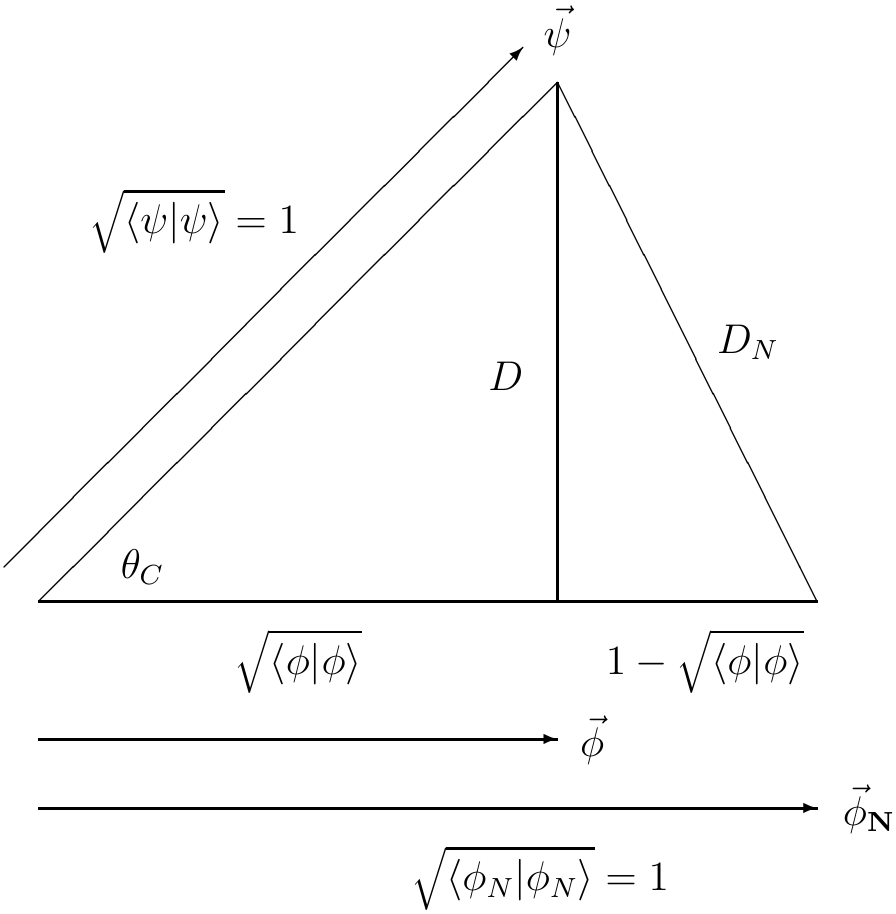}
\caption{Geometrical comparison of the entanglement measure using
unnormalized ($\vec \phi$) and normalized ($\vec \phi_N$) separable states}
\label{geometry}
\end{figure}
\end{center}

The preceding has a close connection to the results of the entanglement
measure using a normalized separable state;
indeed, rewriting Eqs.~(\ref{opt}) for the unnormalized separable
state in terms of variables ${\widetilde a_i} = a_i/\sqrt{N_A}$,
${\widetilde b_j} = b_j/\sqrt{N_B}$, etc. formally leads to the
relations of Eq.~(\ref{normalized}), with the eigenvalue
$\Lambda$ identified with $\sqrt{\iprod{\phi}{\phi}} = \sqrt{N_AN_BN_C\ldots}$. 
This makes for a simple geometrical interpretation of the
results of using unnormalized and normalized separable states
indicated in Fig.~(\ref{geometry}). Both the approach of Eq.~(\ref{normalized}) using
normalized separable states and that of Eq.~(\ref{opt}) using unnormalized states 
lead to the same angle $\cos\theta_C = \Lambda = \sqrt{N_AN_BN_C\ldots}$ of
Eq.~(\ref{cangle}).
The corresponding distances differ, however.
Using unnormalized separable states $\ket{\phi}$,
the distance of Eq.~(\ref{cdistance}) is 
\begin{equation}
D^2 = \iprod{\phi - \psi}{\phi - \psi} = 
1-\left(\sqrt{\iprod{\phi}{\phi}}\right)^2 = 1-\cos^2\theta_C,
\end{equation}
On the other hand, using normalized separable states $\ket{\phi_N}$, the
corresponding distance is  
\begin{equation}
D_N^2 = \iprod{\phi_N - \psi}{\phi_N - \psi} = 
D^2 + \left(1-\sqrt{\iprod{\phi}{\phi}}\right)^2
= 2(1-\cos\theta_C).
\end{equation}
Thus, in the approach using unnormalized separable states,
the use of $1-\cos\theta_C^2 = \sin\theta_C^2$ as the entanglement measure 
can be interpreted as the distance to the closest separable state.
\section{Schmidt decomposition}
As mentioned in the previous section, the equations of (\ref{opt}) 
determining the extremal points of the distance to the closest separable
state are non--linear. One of the special cases for which a closed--form
solution exists is when the $n$--dimensional system is decomposed
into an $u$-dimensional subsystem, $A$, and a $v$-dimensional
subsystem, $B$, such that $n = uv$. In this case the equations
decouple to yield
\begin{eqnarray}
a_k N_AN_B &=& \sum_{i = 0}^{u-1}\sum_{j = 0}^{v-1} \chi_{kj}\chi_{ij}^*a_i 
\nonumber\\
b_m N_AN_B &=& \sum_{j = 0}^{v-1}\sum_{i = 0}^{u-1} \chi_{im}\chi_{ij}^*b_j
\label{extremal} 
\end{eqnarray}
which, respectively, leads to the product $N_AN_B$ being found as solutions to
\begin{eqnarray}
{\rm det} \left| N_AN_B \delta_{ki} - \sum_{j=0}^{v-1} \chi_{kj}\chi_{ij}^*
\right| &=& 0 \nonumber\\
{\rm det} \left| N_AN_B \delta_{mj} - \sum_{i = 0}^{u-1} \chi_{im}\chi_{ij}^*
\right| &=& 0
\label{vn}\end{eqnarray}
\par
This has an interesting correspondence to the Schmidt decomposition of $\ket{\psi}$ of
Eq.~(\ref{sd}). To see this, consider the density matrix of the pure state
$\ket{\psi}$:
\begin{equation}
\rho = \oprod{\psi}{\psi} = \sum_{i=0}^{n-1} \sum_{j=0}^{n-1} \chi_i \chi^*_j \oprod{i}{j}
\end{equation}
where $\chi_i$ are the coordinates of $\psi$ in the computational basis:
$\ket{\psi} = \sum_{i=0}^{n-1}\chi_i\ket{i}$.
We decompose the system into a $u$-dimensional subsystem, $A$, and a $v$-dimensional
subsystem, $B$, such that $n = uv$, and expand the density matrix $\rho$ in the
orthonormal basis
$\oprod{i}{j} = \ket{k}\bra{l} \otimes \ket{q}\bra{r}$,
where each vector is a member of the computational
basis in its respective space. The density matrix then takes the form
\begin{equation}
\rho = \sum_{k=0}^{u-1} \sum_{l=0}^{u-1} \sum_{q=0}^{v-1} \sum_{r=0}^{v-1} 
\chi_{kq} \chi_{lr}^* \oprod{k}{l} \otimes \oprod{q}{r}
\end{equation}
where we have reparameterised the $\chi_i$ coordinates as $\chi_{kq}$.
We now define the partial traces over the two subsystems as
\begin{eqnarray}
\Tr_A(\rho) &=& \sum_{t=0}^{u-1} \left( \bra{t} \otimes \mathds{1}_B\right) \rho
   \left( \ket{t} \otimes \mathds{1}_B\right) \\
\Tr_B(\rho) &=& \sum_{t=0}^{v-1} \left( \mathds{1}_A \otimes \bra{t} \right) \rho
   \left( \mathds{1}_A \otimes \ket{t} \right) 
\end{eqnarray}
where $\mathds{1}_A$ and $\mathds{1}_B$ are the identity matrices in the subspaces of
$A$ and $B$, respectively. The reduced density matrix
$\rho_A$, defined by tracing out over the subsystem $B$, can then be written as
\begin{eqnarray}
\rho_A = \Tr_B(\rho) &=&
\sum_{t=0}^{v-1} \sum_{k=0}^{u-1} \sum_{l=0}^{u-1} \sum_{q=0}^{v-1} \sum_{r=0}^{v-1} 
\chi_{kq} \chi_{lr}^* \oprod{k}{l} \otimes \iprod{t}{q}\iprod{r}{t}
\nonumber\\
&=& \sum_{t=0}^{v-1} \sum_{k=0}^{u-1} \sum_{l=0}^{u-1} 
\chi_{kt} \chi_{lt}^* \oprod{k}{l}
\end{eqnarray}
or, in terms of components,
\begin{equation}
\left(\rho_A\right)_{kl} = \sum_{t=0}^{v-1} \chi_{kt} \chi_{lt}^* 
\label{rhoa}
\end{equation}
Similarly, the reduced density matrix
$\rho_B$, defined by tracing out over the subsystem $A$, can be written as
\begin{equation}
\left(\rho_B\right)_{qr} = \Tr_A(\rho) = \sum_{t=0}^{u-1} \chi_{tq} \chi_{tr}^* 
\label{rhob}
\end{equation}
In terms of the reduced density matrices $\rho_A$ and $\rho_B$ of 
Eqs.(\ref{rhoa}) and (\ref{rhob}), we find that
the extremal conditions of Eqs.(\ref{extremal}) can be written as
\begin{eqnarray}
\sum_{i=0}^{u-1} \left(\rho_A\right)_{ki} a_i &=& N_A N_B a_k\\
\sum_{j=0}^{v-1} \left(\rho_B\right)_{mj} b_j &=& N_A N_B b_m
\end{eqnarray}
Thus, $N_AN_B$ can be interpreted as the eigenvalues of the reduced
density matrices $\rho_A$ and $\rho_B$, with $a_i$ and $b_j$ being
the corresponding eigenvectors. As discussed after Eq.~(\ref{sd}),
this then provides a geometric interpretation of the Schmidt
decomposition of $\ket{\psi}$: the coefficients $a_i$ and $b_j$ used in
defining the closest separable state are related to
the basis states $\ket{\alpha_k}$ and $\ket{\beta_k}$ of the Schmidt decomposition,
with the norm $N_AN_B$ of the closest separable state
related to the Schmidt coefficients $p_k$.
\par
It is interesting to consider
the particular case that the $n$--dimensional space is split into
a product of a single qubit space and another space of dimension
$u=n/2$. In this case,  
one of the equations of (\ref{vn}) will become a quadratic equation
for the product $N_AN_B$, with solutions
\begin{equation}
N_AN_B \equiv \mu_\pm = \frac{1}{2}\left[ 1 \pm \sqrt{1-4C}\right]
, \qquad\qquad C = \sum_{j=1}^{u-1} \sum_{k=0}^{j-1} \left| \chi_{0j}\chi_{1k} - \chi_{1j}\chi_{0k}\right|^2
\end{equation}
The Schmidt decomposition of Eq.~(\ref{sd}) in this case becomes
\begin{equation}
\ket{\psi} = \sqrt{\mu_+} \ \ket{\alpha_+} \otimes \ket{\beta_+} +
 \sqrt{\mu_-} \ \ket{\alpha_-} \otimes \ket{\beta_-}
\end{equation}
with $\mu_+ + \mu_- = 1$.
Relating the eigenvalues $N_AN_B$ to the cosine of the angle
between $\ket{\psi}$ and the closest separable state $\ket{\phi}$ by
Eq.~(\ref{cangle}), and noting that $\mu_+$ is the larger of the
two eigenvalues, we find the Schmidt decomposition can be written as
\begin{equation}
\ket{\psi} = \cos\theta_{\rm max} \ \ket{\alpha_+} \otimes \ket{\beta_+} +
 \sin\theta_{\rm max} \ \ket{\alpha_-} \otimes \ket{\beta_-}
\end{equation}
where $\cos\theta_{\rm max} \equiv \mu_+$. If $\ket{\psi}$ was separable
we would have $\sin^2\theta_{\rm max} = \mu_- = 0$, and as
such this provides a direct connection between the entanglement measures
of $1-\Lambda^2_{\rm max} = \sin^2\theta_C$ in the geometric approach 
and the coefficient $\sin^2\theta_{\rm max}$ in the Schmidt decomposition.
\par
This geometric connection can also be made to the generalization of the
Schmidt decomposition for multipartite pure states developed by
Partovi \cite{s4}. In this approach, one starts with a state $\ket{\psi}$
and decomposes it into two subsystems: one, a qubit space $A$,
and another space ($BC\ldots Z$) representing the remaining dimensions:
\begin{equation}
\ket{\psi} = \sum_{i_a} \sqrt{p_{i_a}^A} \ket{\psi_{i_a}^A} 
\otimes \ket{\psi_{i_a}^{BC\ldots Z}}
\label{stage1}\end{equation}
One then decomposes $ \ket{\psi_{i_a}^{BC\ldots Z}} $ into two
subsystems: another qubit space $B$, and another space
($CD\ldots Z$) representing the remaining dimensions:
\begin{equation}
\ket{\psi_{i_a}^{BC\ldots Z}} = \sum_{i_b} \sqrt{p_{i_a;i_b}^B}
\ket{\psi_{i_a;i_b}^B} \otimes \ket{\psi_{i_a;i_b}^{CD\ldots Z}}
\label{stage2}\end{equation}
This process is continued until the last two qubit spaces
$Y$ and $Z$ are reached, with the result
\begin{equation}
\ket{\psi} = \sum_{i_a i_b \ldots i_{yz} } 
\sqrt{p_{i_a}^A p_{i_a;i_b}^B \cdots 
 p_{i_a i_b \ldots i_x;i_{yz} }^{YZ}  } \ket{\psi_{i_a}^A} 
\otimes \ket{\psi_{i_a;i_b}^B} \otimes\cdots\otimes
 \ket{\psi_{i_a i_b \ldots i_x;i_{yz} }^{Y} } \otimes
 \ket{\psi_{i_a i_b \ldots i_x;i_{yz} }^{Z} }
\end{equation}
There is a direct correspondence between each stage of this series
of decompositions and a problem involving the finding of the extremal
points of a particular distance in the geometric approach described
in the previous section. For example, at the first stage of Eq.~(\ref{stage1}),
we can consider the distance
between $\ket{\psi}$ and
and a state $\ket{\phi^{A;BC\ldots Z}} =  \ket{\phi^A} 
\otimes \ket{\phi^{BC\ldots Z}}$:
\begin{equation}
D_{A;BC\ldots Z}^2 = \iprod { \psi - \left[
\phi^A \otimes \phi^{BC\ldots Z} \right] }
{ \psi - \left[
\phi^A \otimes \phi^{BC\ldots Z} \right] }
\end{equation}
where $\ket{\phi^A}$ is a qubit state and $\ket{\phi^{BC\ldots Z}}$ encompasses
the remaining dimensions. Finding the extremal points of this distance
will result in a system of (linear) equations, as in Eq.~(\ref{vn}),
determining the components of the state $\ket{\phi^{A;BC\ldots Z}}$.
For the next stage, corresponding to Eq.~(\ref{stage2}),
we can then consider the
distance between the state $\ket{\phi^{BC\ldots Z}} $ and a state
$\ket{\phi^{B;CD\ldots Z}} =  \ket{\phi^B} 
\otimes \ket{\phi^{CD\ldots Z}}$:
\begin{equation}
D_{B;CD\ldots Z}^2 = \iprod { \phi^{BC\ldots Z} - \left[
\phi^B \otimes \phi^{CD\ldots Z} \right] }
{ \phi^{BC\ldots Z} - \left[
\phi^B \otimes \phi^{CD\ldots Z} \right] }
\end{equation}
where $\ket{\phi^B}$ is a qubit state and $\ket{\phi^{CD\ldots Z}}$
encompasses the remaining dimensions. Finding the extremal points of this
distance will again result in a system of linear equations
determining the components of the state $\ket{\phi^{B;CD\ldots Z}}$.
This process may be continued until the last two qubit states
$\ket{\phi^Y}$ and $\ket{\phi^Z}$ are reached; at each stage there will
be a direct correspondence between the coefficients used in
defining the closest separable state to
the basis states of the Schmidt decomposition,
with the norm of the closest separable state
related to the corresponding Schmidt coefficients.
At the end, we can then define, in analogy with Eq.(\ref{cangle}),
the cosine of the critical angle $\theta_C$ as
\begin{equation}
\cos\theta_C = \left. \frac{
 \iprod{\psi}{\phi} }
 { \sqrt{\iprod{\phi}{\phi}}\sqrt{\iprod{\psi}{\psi}}}
\right|_{\rm critical} \qquad\qquad \ldots\quad
\ket{\phi} = \ket{\phi^A} \otimes \ket{\phi^B} \otimes \cdots \ket{\phi^Z}
\end{equation}
and then use as a measure of entanglement $\sin^2\theta_C = 1-\cos^2\theta_C$.
Although this procedure has the advantage compared to the approach of
Section II of resulting in a series of linear equations to solve,
compared to the non--linear equations of Eq.~(\ref{opt}), the
disadvantage is that the final
result depends on the order that the series of decompositions is made:
the sequence $ABC\ldots YZ$ described above will differ from the
sequence $BC\ldots ZA$. As such, an approach such as that of Ref.~\cite{s4}
of a minimization over all permutations
of the possible orders of the decompositions must be done.
\section{Conclusions}
We have considered a generalization of the usual
geometric measure of entanglement of pure states using the
distance to the nearest unnormalized product state. Although this doesn't
lead to any computational advantages, as the resulting equations
determining the measure are still non--linear in general, this does afford
an interpretation of the standard entanglement measure $1-\Lambda_{\rm max}^2$
as the distance to the closest separable state. This also
provides a relationship between the the norm and components of the closest separable state
and the coefficients and basis states
of the Schmidt decomposition of the state $\ket{\psi}$.
\acknowledgments
This work was supported by the Natural Sciences and Engineering Research
Council of Canada.


\begin{thebibliography}{99}
\bibitem{nielsen} M.~Nielsen and I.~Chuang, {\it Quantum Computation and Quantum
Information} (Cambridge University Press, 2000).
\bibitem{review1} Ryszard Horodecki, Pawel Horodecki, Michal Horodecki, and Karol Horodecki,
arXiv:quant-ph/0702225 (submitted to Rev.~Mod.~Phys.).
\bibitem{review2} Martin B. Plenio and Shashank Virmani, 
Quant.~Inf.~Comp.~{\bf 7}, 1 (2007).
\bibitem{review3} Karol Zyczkowski and Ingemar Bengtsson, arXiv:quant-ph/0606228v1.
\bibitem{quantify} V.~Vedral, M.~B.~Plenio, M.~A.~Rippin, and P.~L.~Knight,
Phys.~Rev.~Lett.~{\bf 78}, 2275 (1997). 
\bibitem{g1} A.~Shimony, Ann.~N.~Y.~Acad.~Sci.~{\bf 755}, 675 (1995).
\bibitem{g2} H.~Barnum and N.~Linden, J.~Phys.~{\bf A34}, 6787 (2001).
\bibitem{g3} Tzu-Chieh Wei and Paul M.~Goldbart,
Phys.~Rev.~{\bf A68}, 042307 (2003).
\bibitem{g4} Ya Cao and An Min Wang, arXiv:quant-ph/0701099v2.
\bibitem{witness} Tzu-Chieh Wei and Paul M. Goldbart, arXiv:quant-ph/0303079v1
\bibitem{s1} Tsubasa Ichikawa, Izumi Tsutsui, and Taksu Cheon,
arXiv:quant-ph/0702167.
\bibitem{s2} Jon Magne Leinaas, Jan Myrheim, and Eirik Ovrum,
arXiv:quant-ph/0605079.
\bibitem{s3} A.~Yu.~Bogdanov, Yu.~I.~Bogdanov, and K.~A.~Valiev,
arXiv:quant-ph/0512062.
\bibitem{s4} M.~Hossein Partovi, Phys.~Rev.~Lett.~{\bf 92}, 077904 (2004).
\bibitem{s5} Ashish V.~Thapliyal, Phys.~Rev.~{\bf A59}, 3336 (1999).
\end{thebibliography}
\end{document}